\documentclass[prl,twocolumn,aps]{revtex4}
\usepackage{graphicx}

\newcommand{\ket}[1]{\left|#1\right\rangle}

\newcommand{\delom}{\delta\omega}
\newcommand{\nuom}{\omega}

\begin{document}

\title{Ground state  cooling of mechanical resonators}

\author{Ivar Martin$^1$, Alexander Shnirman$^2$, Lin Tian$^{3,4}$, and Peter Zoller$^{3,4}$}

\affiliation{$^1$Theoretical Division, Los Alamos National
Laboratory, Los Alamos, NM 87545, USA\\
$^2$Institut f\"ur Theoretische Festk\"orperphysik,
Universit\"at Karlsruhe, D-76128 Karlsruhe, Germany\\
$^3$ Institute for Theoretical Physics, University of Innsbruck, A-6020, Austria\\
$^4$ Institute for Quantum Optics and Quantum Information of
the Austrian Academy of Sciences, 6020 Innsbruck, Austria
}

\begin{abstract}
We propose an application of a single Cooper pair box (Josephson
qubit) for active cooling of nanomechanical resonators. Latest
experiments with Josephson qubits demonstrated that long coherence
time of the order of microsecond can be achieved in special
symmetry points. Here we show that this level of coherence is
sufficient to perform an analog of the well known in quantum
optics ``laser'' cooling of a nanomechanical resonator
capacitively coupled to the qubit.  By applying an AC driving to
the qubit or the resonator, resonators with frequency of
order 100 MHz and quality factors higher than $10^3$ can be
efficiently cooled down to their ground state, while lower
frequency resonators can be cooled down to micro-Kelvin
temperatures.  We also consider an alternative setup where
DC-voltage-induced Josephson oscillations play the role of the AC
driving and show that cooling is possible in this case as well.
\end{abstract}
\maketitle

\section{Introduction}

Recently, fabrication of nanomechanical resonators with
fundamental frequencies in the microwave range (100 MHz to 1GHz)
has been achieved~\cite{Rouk_1GHz}. For such resonators, the
quantum mechanical level spacing is a few micro-$e$V, which is
comparable to the lowest achievable cryogenic temperatures.
Freezing out the mechanical degrees of freedom is favorable for
ultra sensitive detection applications~\cite{MRFM} due to reduced
effects of thermal fluctuations. Even more spectacular
applications can be envisioned if it is possible to cool the
mechanical systems down to their motional ground states with high
probability. Creation of exotic non-classical states, entanglement
with other quantum  objects, e.g. spins or atoms, coherent quantum
information transfer between quantum sub-systems are just a few
possibilities. However, reaching the motional ground state using
conventional passive cooling techniques is practically unfeasible,
and therefore other approaches need to be explored.  Fortunately,
the cooling problem is not unique to nanomechanics; a similar
problem has been encountered and successfully solved in the field
of ultra-cold atoms, where by using $active$ cooling approaches it
was possible to quench the vibrational motion of atoms and reach
effective nano-Kelvin temperatures~\cite{Leibfried_RMP}.  The
connection to quantum optics has been recently explored in several
works. Hopkins et al.~\cite{Hopkins_Cooling} applied the quantum
feedback control ideas to nanomechanical resonator cooling.
Wilson-Rae et al.~\cite{Wilson_Rae_Cooling} proposed an
analogue of resolved sideband laser cooling by coupling the
resonator displacement to the level spacing of an attached
semiconductor quantum dot, which is being irradiated by
red-detuned laser.  An advantage over the feedback-based
techniques is that the sideband cooling does not require
on-the-fly analysis of the output of a nearly ideal detector. On
the other hand, direct implementation of the
Ref.~\cite{Wilson_Rae_Cooling} approach appears rather
challenging, from the fabrication stand point and due to stringent
constraints on the quantum dot relaxation rate, which needs to be
slower than the resonator frequency for the vibrational sidebands
to be resolved.

In this work we study an alternative realization of the laser-like
cooling for nanomechanical resonators, where the role of the
two-state system is played by a superconducting qubit (Cooper Pair
Box, or CPB) capacitively coupled to the resonator.  Interaction
between the qubit and the resonator leads to splitting of the
qubit states into equidistant vibrational sidebands. Latest
experiments with Josephson qubits showed that long coherence time
of the order of microsecond can be achieved in special symmetry
points~\cite{Saclay_Manipulation_Science}. When the qubit
relaxation and dephasing rates are smaller than the oscillator
frequency, one reaches the resolved sideband regime, favorable for
cooling.  By tuning microwave source frequency into the first red
sideband (qubit level spacing minus the oscillator frequency), one
can ensure that the microwave photon absorption processes are
preferentially accompanied by simultaneous phonon emission from
the resonator. The cooling cycle is completed when a photon is
spontaneously emitted at the qubit natural frequency into an
external bath. The emitted photon is blue-shifted relative to the
source, as it carries away one resonator phonon energy.  The exact
reverse of this process is exponentially suppressed if the qubit
level spacing is large relative to temperature, and hence the
heating is determined by other -- non-resonant or driving-induced
-- processes which are much slower than the dominant cooling
mechanism.  This makes cooling possible. A suggestion to use
coupled nanoresonator-qubit system for laser-like cooling has been
first made by Irish and Schwab~\cite{Irish}.  Here we consider two
implementation of this scheme that use different AC driving
sources: (1) microwaves directly applied to the CPB or the
resonator, in the form of AC flux or voltage; (2) the AC
Josephson effect on an auxiliary Josephson junction of the qubit.
The advantage of the second approach is that it only requires DC
bias for cooling; however, we find that it is not as effective as
the AC biasing scheme.  In the opposite limit of fast CPB
relaxation when the vibrational sidebands are not resolved, we
demonstrate that another type of laser cooling, ``Doppler''
cooling~\cite{Leibfried_RMP}, can be performed down to the
temperature defined by the quantum dot level width. Although
ground state cooling in this regime is impossible, this technique
could be practically attractive for noise reduction in local probe
applications (e.g., AFM, MFM, and MRFM), where typical resonator
frequencies are below 1 MHz.

\section{AC cooling}\label{sec:ac}
\begin{figure}
\includegraphics[width=0.55\columnwidth]{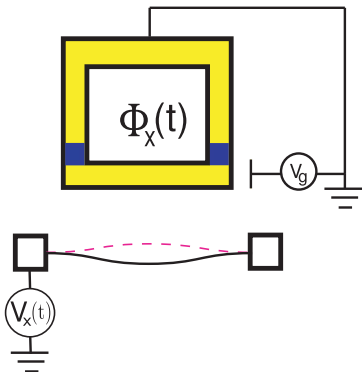}
\caption{The system with AC driving.}
\label{Fig:SQUIDsys}
\end{figure}

{\em System.} The system under consideration is shown in
Fig.~\ref{Fig:SQUIDsys}.  A mechanical resonator (horizontal beam) is coupled
to a Cooper pair box (yellow rectangle) through capacitance $C_x(x)$, which
depends on the resonator displacement $x$.  The charge, and hence the state, of
CPB is separately controlled by the gate voltage $V_g$ applied to capacitor
$C_g$.  CPB is coupled to a large superconductor through two Josephson junctions.
The SQUID geometry is chosen to allow for application of an external AC flux to
the system to provide the AC driving needed for cooling. Similar systems were
considered in Refs.~\cite{ArmourAD:Entadm,Irish}. The Hamiltonian of the system
without dissipation reads
\begin{eqnarray}
\label{Eq:Charge_Hamiltonian_x} H = \frac{(Q-C_{\rm g}V_{\rm
g}-C_x(x)V_x)^2}{2C_{\Sigma}}- E_{\rm J}(\Phi_x)\cos\theta
 + H_x \ ,
\end{eqnarray}
where $Q$ is the charge on the island, $C_{\Sigma}\equiv C +
C_{\rm g} + C_x(x)$ is the total capacitance of the island, and
$H_x \equiv \frac{p^2}{2m}+\frac{m\omega_0^2 x^2}{2}$. The total
Josephson energy of the SQUID controlled by an external flux
$\Phi_x$ is given by $E_{\rm J}(\Phi_x) = 2 E_{\rm J}^{0}\cos(\pi
\Phi_x/\Phi_0)$, where $E_{\rm J}^{0}$ is the Josephson energy of
each of the junctions (we consider a symmetric setup) and $\Phi_0
= h/2e$ is the (superconducting) flux quantum. We assume the total
gate charge $Q_{\rm g}\equiv C_{\rm g}V_{\rm g}+C_x(x)V_x$ to be
close to an odd number of electron charges, i.e., $Q_{\rm g} =
2e(N+1/2) + 2e\delta N$, where $|\delta N| \ll 1/2$. Then we can
use a two state approximation, i.e.,
$\ket{\uparrow}\equiv\ket{Q=2eN}$ and
$\ket{\downarrow}\equiv\ket{Q=2e(N+1)}$ and rewrite the
Hamiltonian using the Pauli matrices as
\begin{eqnarray}
\label{Eq:Spin_Hamiltonian_x} H &=& 4E_{\rm C}(x) \,\delta N(x)\,
\sigma_z + E_{\rm C}(x)(1+4\delta
N^2(x))\nonumber \\
&-& \frac{E_{\rm J}}{2}\,\sigma_x + H_x \ ,
\end{eqnarray}
where $E_{\rm C}(x)\equiv e^2/2C_{\Sigma}$. The second term in the
first line of Eq.~(\ref{Eq:Spin_Hamiltonian_x}) depends on $x$
and, thus, added to the oscillator Hamiltonian $H_x$, renormalizes
(slightly) the oscillator parameters. This term is also
responsible for the direct coupling of the oscillator to the
dissipation in the circuitry (see below). We, first, drop this term for
clarity but later reintroduce it when discussing the direct coupling between
the oscillator and the gate voltage fluctuations. Assuming that
fluctuations of $x$ are small relative to the resonator-CPB
distance $d$, we obtain $C_x(x) \approx C_x - C_x\,x/d$. Thus
$\delta N(x) \approx \delta N - N_x \, x/d$, where $N_x \equiv C_x
V_x/2e$, and $E_C(x) \approx E_C + E_C\,(C_x/C_{\Sigma})\,(x/d)$.
Then the Hamiltonian simplifies to
\begin{eqnarray}
\label{Eq:Spin_Hamiltonian_dx} H = 4E_{\rm C}\, \delta N\,\sigma_z
- \frac{E_{\rm J}}{2}\,\sigma_x + \lambda \,(a^{\dag} +
a)\,\sigma_z + \hbar \omega_0 a^{\dag}a \ ,
\end{eqnarray}
where $\lambda \equiv -4 E_{\rm C}\,\left[N_x - \delta N
(C_x/C_{\Sigma})\right]\,(\Delta x/d)$ and $\Delta x \equiv
\sqrt{\hbar/(2m\omega_0)}$ is the amplitude of zero point motion,
$x=\Delta x (a^{\dag}+a)$. To increase the coupling one usually
applies high gate voltage $V_x$ so that $N_x \gg 1$. Thus,
approximately
\begin{equation}
\label{Eq:lambda} \lambda \approx  -4 E_{\rm C} N_x {\Delta x\over
d}\ .
\end{equation}

Both gate voltages, $V_g$ and $V_x$, as well as the external flux
$\Phi_x$ fluctuate as they are provided by dissipative sources.
This makes $\delta n$ and $E_{\rm J}$ in
Eq.~(\ref{Eq:Spin_Hamiltonian_dx}) fluctuate. Moreover, in all
real systems there are $1/f$ charge and flux noises. The charge
$1/f$ noise can effectively be added to the noise of the gate
charge $\delta N$. Experimentally, $1/f$ noise is the most severe
factor limiting coherence. Long coherence times have been achieved
in Ref.~\cite{Saclay_Manipulation_Science} by operating in a
special point where the $1/f$ noise is less harmful. In the
special point, the total energy splitting of the qubit $\Delta E
\equiv \sqrt{(8 E_{\rm C})^2\, \delta N^2 + E_{\rm J}^2}$ is not
sensitive to the fluctuations of $\delta N$ and $\Phi_x$ in the
linear order. This implies $\langle \delta N \rangle=0$ and
$\partial E_{\rm J}/\partial \Phi_x =0$. The fluctuations of
$\delta N$ are, nevertheless, still there and we obtain the
following Hamiltonian characterizing the special point
\begin{eqnarray}
\label{Eq:Spin_Hamiltonian_X} H =&-& \frac{E_{\rm J}}{2}\,\sigma_x
+ \hbar \omega_0 a^{\dag}a + \lambda
\,(a^{\dag} + a)\,\sigma_z \nonumber \\&-& \frac{X}{2}\, \sigma_z +
g\,(a^{\dag} + a)\, X\ ,
\end{eqnarray}
where $X \equiv -\frac{4 E_{\rm C}}{e}\,\left(C_{\rm g} \delta
V_{\rm g} + C_x \delta V_x \right)$.
By construction the coupling constant $\lambda$ also contains a fluctuating
part. We neglect this higher order effect.
The fluctuations of voltages
$\delta V_{\rm g}$ and $\delta V_x$ are caused by external
impedances $Z(\omega)$ of the circuits that supply the voltages.
The (unsymmetrized) noise power is given by $\langle \delta
V^2_{\omega} \rangle \equiv \langle \delta V(t) \delta V(0)
\rangle_{\omega} = [{\rm Re} Z(\omega)]\hbar \omega
(\coth(\hbar\omega/2k_{\rm B}T)+1)$. Thus the term $X\sigma_z/2$
represents the coupling of the system to the harmonic (Gaussian)
electromagnetic bath.  The situation is similar to the quantum
optics one where an atom is coupled to electromagnetic vacuum. The
main differences are that in our case the coupling is
longitudinal, i.e., it does not cause spin flips in the natural
(charge) basis, and the bath temperature cannot always be
neglected. The last term in Eq.~(\ref{Eq:Spin_Hamiltonian_X}) is
the direct coupling between the bath and the oscillator.  It
originates from the second term in
Eq.~(\ref{Eq:Spin_Hamiltonian_x}), and $g = N_x\Delta x/d$ at the symmetry point.
In the weak coupling case that we consider, the bath is fully
characterized by its spectral function, and hence we do not need
to include the self-Hamiltonian of the bath explicitly.   The
non-electromagnetic bath acting on the resonator is later
introduced through a phenomenological quality factor $Q$.

{\em Equilibrium transition rates.} To proceed, we first need to determine the
resonator-qubit system relaxation rates caused by the electromagnetic
environment.  It is convenient to perform a $\pi/2$ rotation in the $x-z$
plane, $\sigma_x \leftrightarrow \sigma_z$, to the eigenbasis of the qubit at
the symmetry point. The Hamiltonian becomes
\begin{eqnarray}
\label{Eq:Spin_Hamiltonian_rotated} H =&-& \frac{E_{\rm
J}}{2}\,\sigma_z + \hbar \omega_0
a^{\dag}a + \lambda \,(a^{\dag} + a)\,\sigma_x
\nonumber \\  &-& \frac{X}{2}\, \sigma_x +
g\,(a^{\dag} + a)\, X
\ .
\end{eqnarray}
We choose $E_{\rm J} \gg \omega_0,\lambda,T$. This ensures that without the
driving the spin is all the time in the ground state $\ket{\uparrow}$. The
qubit relaxation rate is
\begin{equation}
\label{Eq:Gamma_r}
\Gamma_{\rm r} \equiv \Gamma_{\downarrow \rightarrow \uparrow} =
\frac{\langle X^2_{\nuom=E_{\rm J}}\rangle}{4\hbar^2}
\approx \pi\alpha_g E_{\rm J}
\ ,
\end{equation}
where $\langle X^2_{\nuom}\rangle = 2\pi\hbar^2 \alpha_g \nuom
[\coth(\hbar\nuom/2k_{\rm B}T)+1]$,
$\alpha_g \approx \frac{C_x^2+C_{\rm g}^2}
{C_{\Sigma}^2}\,\frac{R}{R_Q}$, $R_Q\equiv h/4e^2$ (here we assumed independent
$\delta V_{\rm g}$ and $\delta V_x$ with similar external impedances
${\rm Re} Z(\omega) \equiv R\sim 50\ \Omega$).
The opposite excitation rate is exponentially suppressed.

\begin{figure}
\includegraphics[width=0.55\columnwidth]{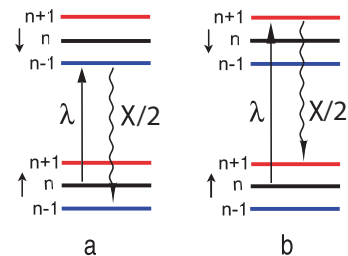}
\caption{Dissipative processes due to the presence of the qubit:
a) $n\rightarrow n-1$; b) $n\rightarrow n+1$. The spectra of the
oscillator and the qubit are superimposed.} \label{Fig:Add_Diss}
\end{figure}

The direct coupling between the bath and the oscillator gives the dissipative
rates between the oscillator states $\ket{n}$: $\Gamma_{n \rightarrow n-1}
\approx g^2 \langle X^2_{\nuom=\omega_0}\rangle n /\hbar^2$ and $\Gamma_{n
\rightarrow n+1} \approx g^2 \langle X^2_{\nuom=-\omega_0}\rangle (n+1) /\hbar^2$. In addition,
the oscillator can relax via the virtual excitations of the qubit. The
corresponding processes are shown in Fig.~\ref{Fig:Add_Diss}.
The rates for these processes are given by
\begin{equation}
\Gamma_{n \rightarrow n-1} \approx \frac{\lambda^2}{E_{\rm J}^2}\,
\frac{\langle X^2_{\nuom=\omega_0}\rangle}{4\hbar^2}\,n
\ ,
\end{equation}
\begin{equation}
\Gamma_{n \rightarrow n+1} \approx \frac{\lambda^2}{E_{\rm J}^2}\,
\frac{\langle X^2_{\nuom=-\omega_0}\rangle}{4\hbar^2}\,(n+1)
\ .
\end{equation}
These rates are bigger than those due to the direct coupling by a factor $(2
E_{\rm C}/E_{\rm J})^2$, and hence the direct coupling term in the Hamiltonian
($\propto g$) can be discarded.  The above transition rates translate into the
resonator quality factor caused by the coupling to the electromagnetic bath,
\begin{eqnarray}
\label{Eq:Qfactor}
{1\over Q_{\rm em}} = \frac{\pi \lambda^2}{E_{\rm J}^2}\alpha_g.
\end{eqnarray}

The charge noise at the (relatively low) frequency $\nuom=\omega_0$ may be
dominated by the $1/f$ contribution. The symmetrized correlator of this
contribution has been studied, e.g., in echo experiments~\cite{Nakamura_Echo},
and one can assume $S_X(\nuom)=(\langle X_{\nuom}^2\rangle + \langle
X_{-\nuom}^2\rangle)/2 = E^2_{1/f}/|\nuom|$, where $E_{1/f} \equiv 4E_C
\sqrt{\alpha_{1/f}}$ and typically $\alpha_{1/f} \approx 10^{-7}$. We are not
aware of any study of the unsymmetrized correlators. Thus, we will introduce a
phenomenological temperature $T_{1/f}$ for the particular frequency
$\nuom=\omega_0$ via $\exp\left[-\hbar\omega_0/(k_{\rm B}T_{1/f})\right] \equiv
\langle X_{-\omega_0}^2\rangle/\langle X_{\omega_0}^2\rangle$. Then, for the corresponding
quality factor we obtain
\begin{equation}
{1\over Q_{1/f}} \approx \frac{\lambda^2}{E_{\rm J}^2}
\,\frac{8 E_C^2}{(\hbar\omega_0)^2}\,\alpha_{1/f}\tanh{\hbar\omega_0\over{2k_{\rm B}T_{1/f}}}
\ .
\end{equation}
It is reasonable to assume that the effective $1/f$ noise temperature is
not lower than the environment temperature. For further analysis
we will assume $T_{1/f} \approx T$.

\subsection{Cooling by applying flux driving}

For cooling we need to drive the system out of equilibrium. We
propose to apply an external AC flux, $\Phi_x(t) = \Phi_{x,0} +
D\Phi_0\cos\omega_{\rm d}\, t$, to the qubit. As we operate at the special point where
$\partial E_{\rm J}/\partial \Phi_x =0$, linear driving is impossible. In the
quadratic order, we obtain $E_{\rm J} \rightarrow E_{\rm J} + (\pi D/2)^2E_{\rm
J}(\cos 2\omega_{\rm d} t + 1)$. This gives
\begin{eqnarray}
\label{Eq:Spin_Hamiltonian_pumped} H =&-& \frac{E_{\rm
J}+\Omega}{2}\,\sigma_z - \frac{X}{2}\, \sigma_x + \hbar \omega_0
a^{\dag}a \nonumber \\&+& \lambda \,(a^{\dag} + a)\,\sigma_x -
\frac{\Omega}{2}\, \sigma_z \cos 2\omega_{\rm d}\,t \, ,
\end{eqnarray}
where $\Omega = (\pi D/2)^2E_{\rm J}$.  To describe the cooling process we use
the Floquet picture. In other words, we count the number of energy quanta taken
from the pumping source. This amounts to substituting the factors $e^{\pm
2\omega_{\rm d}\,t}$ by raising/lowering operators $e^{\pm \chi}$ ($e^{i\chi}|m\rangle =
|m+1\rangle$) and subtracting the energy taken from the source, $2\hbar\omega
m$, from the Hamiltonian. Then the new Hilbert space of the problem is extended
as $|\sigma\rangle|n\rangle|m\rangle$, where
$|\sigma\rangle$ is the state of the qubit ($\ket{\uparrow}$ or
$\ket{\downarrow}$), while $n$ is the number of quanta in the resonator. Due to
the new term in the Hamiltonian, $-2\hbar\omega_{\rm d}m$, an unbound staircase of
states appears. Each step of the staircase (Floquet zone) is characterized by a
number $m$ and spans the Hilbert space $\ket{\sigma}\ket{n}$. The neighboring
zones are shifted by energy $2\hbar
\omega_{\rm d}$ relative to each other. The states $\ket{\sigma}\ket{n}\ket{m}$ are the
eigenstates of the Hamiltonian
\begin{eqnarray}
H_0 \equiv - (E_{\rm J}+\Omega)\,\sigma_z/2\ + \hbar \omega_0 a^{\dag}a
-2\hbar\omega_{\rm d}m\,.
\end{eqnarray}
The other three terms form the perturbation
\begin{eqnarray}
H^{\prime} =  - \frac{X}{2}\, \sigma_x + \lambda \,(a^{\dag} + a)\,\sigma_x -
\frac{\Omega}{2}\, \sigma_z \cos \chi
\ .
\end{eqnarray}
This perturbation $H^{\prime}$ causes transitions inside of each zone and also
down and up the staircase. At finite (non-infinite) temperature of the bath the
down transitions prevail and the system propagates down the staircase. This
corresponds to the flow of energy from the driving source to the bath. During
this flow, the driving source quanta of energy can be up- or down-converted in
frequency by amount $\omega_0$, that is, photons with frequency $2\omega_{\rm d}\pm
\omega_0$ are emitted into the bath. The former case corresponds to cooling
while the latter to heating.

\begin{figure}
\includegraphics[width=0.55\columnwidth]{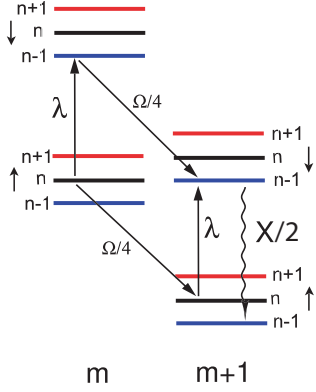}
\caption{The cooling process.} \label{Fig:Stair_Cooling}
\end{figure}

\begin{figure}
\includegraphics[width=0.55\columnwidth]{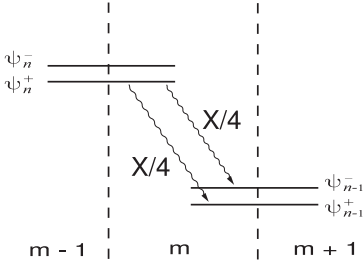}
\caption{Level structure of the strong driving for ac cooling.
Here $\psi_n^{\pm}\equiv (\ket{n,\uparrow}\pm
\ket{n-1,\downarrow})/\sqrt{2}$.}
\label{Fig:Strong_Driving_Cooling}
\end{figure}

We first consider the up-conversion cooling process shown in
Fig.~\ref{Fig:Stair_Cooling}.
\begin{figure}
\includegraphics[width=0.55\columnwidth]{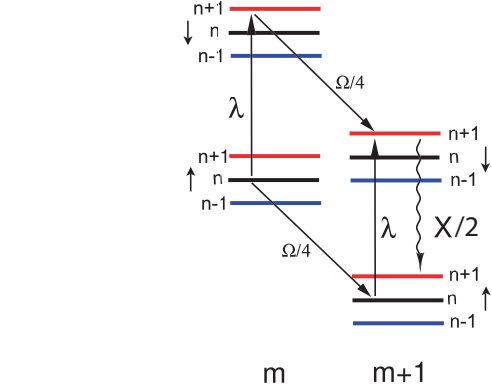}
\caption{The heating process.} \label{Fig:Stair_Heating}
\end{figure}
We choose the optimal detuning so that $2\omega_{\rm d}= E_{\rm J}+\Omega
- \omega_0$ and the levels $\ket{\uparrow,n,m}$ and
$\ket{\downarrow,n-1,m+1}$ are degenerate (in what follows we use the units $\hbar=1$ for
brevity). Thus we have to be
careful when calculating the rates. First, we obtain the second
order matrix element between these two states equal to
$\Delta\,\sqrt{n}$ where
\begin{equation}
\label{Eq:Delta} \Delta \equiv \frac{\Omega\lambda}{2 E_{\rm J}} \
.
\end{equation}
Note that the two paths shown in Fig.~\ref{Fig:Stair_Cooling}
interfere constructively. Then, the cooling rate depends on the
relation between $\Delta\,\sqrt{n}$ and $\Gamma_{\rm r}/2$. For
the {\it weak driving} case $\Delta\,\sqrt{n} < \Gamma_{\rm r}/2$
we have the cooling rate $\Gamma_{n \rightarrow n-1}^{\Omega}
\approx (4\Delta^2/\Gamma_{\rm r})\,n$ (the superscript $\Omega$
is to emphasize that this rate is due to the driving). To justify
this result we consider this process as tunneling from the level
$\ket{\uparrow,n,m}$ to the broadened level
$\ket{\downarrow,n-1,m+1}$. The retarded Green's function of the
second level is then given by $G^{\rm R}(\omega) = 1/(\omega +
i\Gamma_{\rm r}/2)$. Using the Golden rule (the adiabatic
elimination technique\cite{laser_cooling_1992}), we obtain
\begin{eqnarray}
\label{Eq:Golden_Rule}
\Gamma_{n \rightarrow n-1}^{\Omega} &=&
2\pi\,\Delta^2\,n \left(-\frac{1}{\pi}\,{\rm Im}\,G^{\rm R}(\omega=0)
\right) \nonumber \\
&=& \frac{4\Delta^2}{\Gamma_{\rm r}}\,n
\ .
\end{eqnarray}
In the opposite, {\it strong driving} case, $\Delta\sqrt{n} >
\Gamma_{\rm r}/2$, there are coherent oscillations between the two
levels. The appropriate description is then to say that doublets
of new eigenstates, are formed which are split in energy by
$2\Delta \sqrt{n}$, see Fig.~\ref{Fig:Strong_Driving_Cooling}. The
doublets are defined for $n\ge 1$ as $\psi_n^{\pm}\equiv
(\ket{n,\uparrow}\pm \ket{n-1,\downarrow})/\sqrt{2}$ and for $n=0$
we have a single state $\psi_{n=0}\equiv \ket{0,\uparrow}$.
There are four possible transitions between
each state of the doublet and each state of another
doublet shifted as $n\rightarrow n-1$ and $m\rightarrow m+1$.
All four rates are equal
$\Gamma_{\rm r}/4$. In total this gives
$\Gamma_{n \rightarrow n-1}^{\Omega} \approx \Gamma_{\rm r}/2$.
To summarize,
\begin{eqnarray}
\label{Eq:Cooling_Rates}
\Gamma_{n \rightarrow n-1}^{\Omega} =
\left\{\begin{array}{c l} (4\Delta^2/\Gamma_{\rm r})\,n
& {\rm\quad if \quad}\Delta\sqrt{n} < \Gamma_{\rm r}/2 \\
\Gamma_{\rm r}/2  & {\rm\quad if \quad}\Delta\sqrt{n} > \Gamma_{\rm r}/2
\end{array}\right.
\ .
\end{eqnarray}
Note that in the strong driving regime $n$ denotes the doublet rather than
the oscillator's level (see discussion below).

\begin{figure}
\includegraphics[width=0.55\columnwidth]{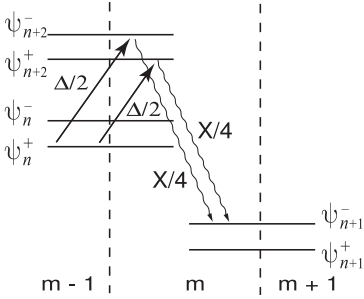}
\caption{Level structure of the strong driving for ac heating. The
two shown processes interfere constructively.}
\label{Fig:Strong_Driving_Heating}
\end{figure}

In addition to cooling, the AC driving induces competing heating
processes. One, shown in Fig.~\ref{Fig:Stair_Heating}, is a standard in
quantum optics off-resonance process. Indeed, the states
$\ket{\uparrow,n,m}$ and $\ket{\downarrow,n+1,m+1}$ are connected
in the second order by the matrix element $\Delta\,\sqrt{n+1}$.
These states are off-resonance (we restrict ourselves to the values of
$n$ such that level splitting $2\omega_0
\gg \Delta\,\sqrt{n+1}$). Thus, in the case $\Delta\,\sqrt{n} <
\Gamma_{\rm r}/2$ we obtain $\Gamma_{n \rightarrow n+1}^{\Omega}
\approx (\Delta/2\omega_0)^2\,\Gamma_{\rm r}\,(n+1)$. In the
strong driving case $\Delta\,\sqrt{n} > \Gamma_{\rm r}/2$ we use
again the basis of doublets, Fig.~\ref{Fig:Strong_Driving_Heating},
and arrive at $\Gamma_{n \rightarrow
n+1}^{\Omega} \approx (\Delta/4\omega_0)^2\,\Gamma_{\rm
r}\,(n+1)\,(1+\delta_{n,0})$.
In this case the rates are between
doublets as a whole. The factor $(1+\delta_{n,0})$ is due to the
fact that there is only one state in the ``doublet'' $n=0$. To
summarize,
\begin{equation}
\label{Eq:Heating_Rates}
\Gamma_{n \rightarrow n+1}^{\Omega} = \frac{\Delta^2}{4\omega_0^2}\,\Gamma_{\rm r}\,(n+1)
\left\{\begin{array}{c l} 1
& {\rm\quad if \ }\Delta\sqrt{n} < \Gamma_{\rm r}/2 \\
\frac{1+\delta_{n,0}}{2}
& {\rm\quad if \ }\Delta\sqrt{n} > \Gamma_{\rm r}/2
\end{array}\right.
\ .
\end{equation}
\begin{figure}
\includegraphics[width=0.55\columnwidth]{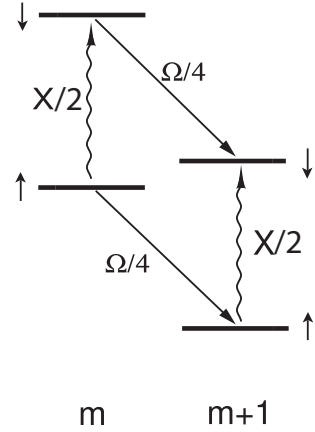}
\caption{Qubit heating induced by the applied drive.}
\label{Fig:Spin_Heating}
\end{figure}

Another driving-induced heating process, not characteristic for quantum optics,
is due to the fact that in solid state systems there is strong noise at low
frequencies ($1/f$ noise). Thus, processes like the one shown in
Fig.~\ref{Fig:Spin_Heating} become relevant. This process excites the qubit
with the rate
\begin{equation}
\label{Eq:Qubit_DI_Excitation}
\Gamma_{\rm e}\equiv
\Gamma_{\uparrow \rightarrow \downarrow}=\frac{\Omega^2}{4 E^2_{\rm J}}\,
\frac{\langle X^2_{\nuom=-\omega_0}\rangle}{4\hbar^2}
\ .
\end{equation}
Since at $\nuom=-\omega_0$ the noise is dominated by the $1/f$ contribution, rate
(\ref{Eq:Qubit_DI_Excitation}) might dominate the heating. With this process,
the heating is also resonant and, in analogy to Eq.~(\ref{Eq:Cooling_Rates}),
still assuming that $\Gamma_{\rm e} \ll
\Gamma_{\rm r}$, we obtain
\begin{eqnarray}
\label{Eq:Heating_Rates_1/f}
\Gamma_{n-1 \rightarrow n}^{\Omega} =
\left\{\begin{array}{c l} \Gamma_{\rm e}\,(4\Delta^2/\Gamma^2_{\rm r})\,n
& {\rm\quad if \quad}\Delta\sqrt{n} < \Gamma_{\rm r}/2\\
\frac{\Gamma_{\rm e}}{2}\,(1+\delta_{n,0})    & {\rm\quad if \quad}\Delta\sqrt{n} > \Gamma_{\rm r}/2
\end{array}\right.
\ .
\nonumber\\
\end{eqnarray}

{\em Master equation.} Taking into account the internal
dissipation of the resonator with quality factor $Q$, one
can~\cite{Wilson_Rae_Cooling} write down the master equation for
the probability $P_n$ to find the resonator in the state
$\ket{n}$. First, we analyze the (strong driving) limit, $\Delta >
\Gamma_{\rm r}/2$. Then, it is convenient to write the rate
equation in terms of the probabilities $D_n$ to find the system in
the doublet $\psi_n^{\pm}$ ($n\ge 1$). The probability $D_0$,
then, is the for the system to be in the (non-degenerate) ground
state $\psi_0=\ket{\uparrow,0}$. The probabilities $P_n$ are given
by
\begin{equation}
P_n=\frac{1}{2}\,(D_n + D_{n+1}) + \frac{1}{2}\,D_n \delta_{n,0}
\ .
\end{equation}
The master equation reads
\begin{eqnarray}
\label{Eq:ME_Strong_Driving}
\dot D_n &=& \frac{1}{2}\,\Gamma_{\rm r}\,\left[D_{n+1}-D_n(1-\delta_{n,0})\right]
\nonumber \\
&+&\frac{\Delta^2}{8\omega_0^2}\;\Gamma_{r}\; \left[nD_{n-1}\left(
1+\delta _{n,1}\right) - \left( n+1\right) D_{n}\left( 1+\delta
_{n,0}\right)\right]
\nonumber \\
&+& \frac{1}{2}\,\Gamma_{\rm
e}\,\left[D_{n-1}(1-\delta_{n,0})(1+\delta_{n,1})-D_n\,(1+\delta_{n,0})\right]
\nonumber \\
&+&
\frac{\omega_0(n_B(\omega_0)+1)}{Q}\times\nonumber \\
&&\left[(n+\frac{1}{2})D_{n+1} - (n-\frac{1}{2})
D_n(1-\delta_{n,0})\right]\nonumber \\
&+&\frac{\omega_0 n_B(\omega_0)}{Q}
\left[\left(n-\frac{1}{2}\right)D_{n-1}(1-\delta_{n,0})(1+\delta_{n,1})\right.
\nonumber \\ &-&\left. \left(n+\frac{1}{2}\right)
D_n\,(1+\delta_{n,0})\right] \ .
\end{eqnarray}
The unusual form of the second part of this master equation is due
to the structure of the matrix elements between the different
states of the doublets.

Multiplying Eqs.~(\ref{Eq:ME_Strong_Driving}) by and summing over
$n$ we obtain
\begin{eqnarray}
\label{Eq:dndt} &&\frac{d}{dt} \langle \tilde n\rangle = \nonumber
\\ &-&\frac{1}{2}\,\Gamma_{\rm r}(1-D_0) + \frac{1}{2}\,\Gamma_{\rm
e}(1+D_0) + \frac{\Delta^2}{8\omega_0^2}\;\Gamma_{\rm r}(1+D_0)
\nonumber \\
&-&\left(\frac{\omega_0}{Q}-\frac{\Delta^2}{8\omega_0^2}\;\Gamma_{\rm
r}\right) \langle \tilde n\rangle +\frac{\omega_0}{2Q}\,(1-D_0) +
\frac{\omega_0\,n_B(\omega_0)}{Q}
\ ,\nonumber \\
\end{eqnarray}
where $\langle \tilde n\rangle \equiv \sum_n n D_n$.
We rewrite the RHS of Eq.~(\ref{Eq:dndt}) in terms of
the phonon occupation expectation value
\begin{eqnarray}
\label{Eq:n_tilde_n}
\langle n \rangle = \sum_{n=1}^\infty{n P_n} =
\sum_{n=1}^\infty{(n-\frac12)D_n} =
\langle \tilde n\rangle - \frac{1-D_0}2
\ .
\end{eqnarray}
We, then, obtain
\begin{eqnarray}
\label{Eq:dn_real_dt}
&&\frac{d}{dt} \langle \tilde n\rangle =\nonumber \\
&-&\frac{1}{2}\,\Gamma_{\rm r}(1-D_0) + \frac{1}{2}\,\Gamma_{\rm
e}(1+D_0)+\frac{\Delta^2}{16\omega_0^2}\;\Gamma_{\rm r}(3+D_0)
\nonumber \\
&-& \left(\frac{\omega_0}{Q}-\frac{\Delta^2}{8\omega_0^2}\;
\Gamma_{\rm r}\right) \langle  n\rangle +\frac{\omega_0}{Q}
n_B(\omega_0) \ .\nonumber \\
\end{eqnarray}
There are two main cooling regimes in (\ref{Eq:dn_real_dt}).
If the coefficient in front of $\langle n \rangle$ is positive,
i.e., if $\omega_0/Q \gg (\Delta^2/8\omega_0^2)\,\Gamma_{\rm r}$,
we obtain the usual cooling with the cooling rate slowing down with
decreasing $\langle n\rangle$. As this is the
relevant regime for realistic parameters we will analyze only this
case. Interestingly, however, in the opposite case of very high
$Q$, when $\omega_0/Q \ll (\Delta^2/8\omega_0^2)\,\Gamma_{\rm r}$,
the cooling rate accelerates until $\langle n\rangle\approx 1$.

If $\langle \tilde n \rangle \gg 1$ (to be checked for
self-consistency) the probability to be in the ground state is
negligible, $D_0 \ll 1$. Then
\begin{equation}
\label{Eq:n_high}
\langle n\rangle \approx n_B(\omega_0) - \frac{(\Gamma_{\rm r}-\Gamma_{\rm e}) Q}
{2\omega_0}
\ .
\end{equation}
This regime, thus, is realized when $n_B(\omega_0) >
(\Gamma_{\rm r}-\Gamma_{\rm e}) Q/(2\omega_0)$.
Clearly, for at least some cooling we need $(\Gamma_{\rm r}-\Gamma_{\rm e}) \gg (2\omega_0)/Q$.
At lower temperatures, Eq.~(\ref{Eq:n_high}) gives negative
$\langle n \rangle$
which means that the approximation breaks down and
$\langle n \rangle \sim 1$
or less. Then, for an estimate we can use $1-D_0 \sim
\langle \tilde n \rangle$. In this regime $\langle \tilde n \rangle \approx 2\langle n \rangle$
(see Eq.(\ref{Eq:n_tilde_n})), therefore $1-D_0 \sim 2\langle n \rangle$.
Thus, for $\Gamma_{\rm r}\gg \Gamma_{\rm e}$
\begin{equation}
\langle n\rangle \approx \frac{\Gamma_{\rm e} +
\frac{\omega_0 n_B(\omega_0)}{Q}}{\Gamma_{\rm r}} \ .
\end{equation}
For $n_B(\omega_0) > \Gamma_{\rm e}Q/\omega_0$ we then obtain
\begin{equation}
\label{Eq:n_medium}
\langle n\rangle \approx \frac{\omega_0 n_B(\omega_0)}{Q\Gamma_{\rm r}}
\ ,
\end{equation}
while in the opposite case the average occupation saturates at
\begin{equation}
\label{Eq:n_low}
\langle n\rangle\approx \frac{\Gamma_{\rm e}}{\Gamma_{\rm r}}
\ .
\end{equation}
From Eq.~(\ref{Eq:dndt}) it is clear that for $n_B(\omega_0)\gg 1$,
the cooling is initially exponential in time, $\langle n\rangle_t \approx
n_B(\omega_0) - \Gamma_{\rm r}Q/(2\omega_0)
\left[1-\exp(-\gamma t)\right]$, with the decay rate determined by
the oscillator bare damping, $\gamma = \omega_0/Q$.
Only when the low occupancy regime
($\langle n\rangle_t < 1 $) is reached,
the rate of exponential decay increases to
\begin{equation}
\gamma = \frac{\Gamma_{\rm r}-\Gamma_{\rm e}}{2} -\frac{\omega_0}{2Q}
\ ,
\end{equation}

If the driving is weaker, then for some low values of $n$ the following master
equation holds (see Refs.~\cite{Wilson_Rae_Cooling,Leibfried_RMP})
\begin{eqnarray}
\label{Eq:ME_Weak_Driving}
\dot P_n &=&\left(A_{-} +
\frac{\omega_0(n_B(\omega_0)+1)}{Q}\right)\left[(n+1)P_{n+1} - n
P_n\right]\nonumber \\
&+&\left(A_{+} + \frac{\omega_0 n_B(\omega_0)}{Q}\right)\left[nP_{n-1} -
(n+1) P_n\right]
\ ,
\end{eqnarray}
where $A_{-} \equiv 4\Delta^2/\Gamma_{\rm r}$ and $A_{+} \equiv
(\Delta/2\omega_0)^2\,\Gamma_{\rm r} + \Gamma_{\rm
e}\,(4\Delta^2/\Gamma^2_{\rm r})$. For the average occupation
number $\langle n\rangle$ in the stationary state we obtain
\begin{equation}
\label{Eq:n_weak}
\langle n\rangle = \frac{A_{+} + \frac{\omega_0 n_B(\omega_0)}{Q}}
{A_{-}-A_{+}+
\frac{\omega_0}{Q}} \ .
\end{equation}
Regimes similar to Eqs.~(\ref{Eq:n_medium}) and (\ref{Eq:n_low}) are clearly
identified. However, a regime similar to Eq.~(\ref{Eq:n_high}) is not possible
within master equation (\ref{Eq:ME_Weak_Driving}). The expression for the
oscillator occupancy Eq.~(\ref{Eq:n_weak}) can be naturally interpreted in
terms of two independent heat baths acting on the resonator, one being the
equilibrium environment at the nominal external temperature $T$, coupled to the
resonator by a coupling strength $\gamma_0 = \omega_0/Q$, while the other bath
being introduced by the cooling process itself.   The effective temperature of
the latter, $T^*$, can be defined through $A_-/A_+ = \exp(\omega_0/T^*)$, and
the effective coupling strength is $\gamma^* = \omega_0/Q^* = A_- - A_+$.
Then, the final resonator occupancy can be re-expressed as
\begin{equation}
\label{Eq:n_2baths}
\langle n\rangle = {\gamma_0 n_B(\omega_0) +
\gamma^* n^*_B(\omega_0)\over \gamma_0 + \gamma^*},
\end{equation}
where $n_B^*$ is the Bose distribution function at temperature $T^*$. Clearly,
$T^*$ is the lowest possible temperature for a given cooling process, which is
achieved for $\gamma^* \gg \gamma_0$. The combined damping coefficient $\gamma
=\gamma_0 + \gamma^*$ determines the rate of relaxation to the new stationary
state.

{\em Example.} We consider a nanomechanical resonator with fundamental
frequency 100 MHz ($\omega_0 = 2\pi \times 100\ {\rm MHz}= 0.5\ \mu$eV) and
quality factor $Q=10^5$. It is coupled to the qubit, which is characterized by
the Josephson energy $E_{\rm J}\approx 50\ \mu$eV and Coulomb charging energy
$E_{\rm C} \approx 160\ \mu$eV (corresponds to $C_{\Sigma}\approx 500$ aF).
The coupling strength between the resonator and the qubit is determined by the
mutual capacitance $C_x\approx 20$~aF, and the gate voltage $V_x \approx 1$ V,
such that $n_x \approx 60$ Cooper pairs (see Ref.~\cite{ArmourAD:Entadm}). The
gap between the resonator and the CPB is $d\approx 100$ nm. For these
parameters, from Eq.~{\ref{Eq:lambda}}, the resonator-CPB coupling strength is
$\lambda \approx 0.1\ \mu$eV.  Assuming that $C_g < C_x$, for the relaxation
rate of CPB we find $\Gamma_{\rm r} = 3\cdot10^{-3}\ \mu$eV.

The circuit-induced quality factor of the resonator is $Q_{\rm em}\approx
4\cdot 10^9$, which is significantly higher than the quality factors of typical
resonators. For the $1/f$ contribution at $T_{1/f} > 10\,\omega_0$ we obtain
$Q_{1/f} > 10^7$, which is still higher than a typical value. Hence this
modification of the oscillator damping can be neglected compared to other
environmental effects.

\begin{figure}
\includegraphics[width=0.9\columnwidth]{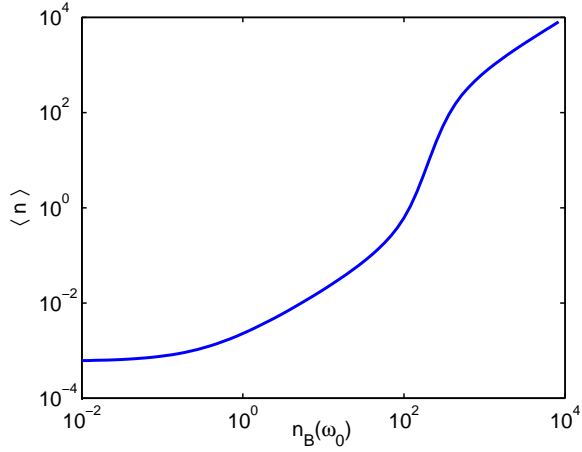}
\caption{Cooling diagram for a 100 MHz resonator (detailed
parameters in text).} \label{Fig:Cool}
\end{figure}

For cooling, we take $\Omega \approx 1\ \mu$eV, which corresponds
to the modulation depth $D = 0.1$. We thus obtain $\Delta \approx
10^{-3}\, \mu$eV, and hence for all $n$ we have $\Delta\,\sqrt{n}
> \Gamma/2$ and $\Gamma_{n \rightarrow n-1}^{\Omega} \approx
\Gamma_{\rm r}/2$ (strong driving regime). The heating is indeed
dominated by $1/f$ noise, with $\Gamma_{\rm e}=(\Omega/2E_{\rm
J})^2\,(E_{1/f}^2/4\omega_0) \approx 1.8\cdot10^{-6}\ \mu$eV.  In
this regime we get the following results for cooling:
\begin{eqnarray}
\langle n\rangle = \left\{\begin{array}{c l} n_B(\omega_0) - 300
& {\rm\quad if \quad}n_B(\omega_0) > 300 \\
1.5 \cdot 10^{-3}n_B(\omega_0)  & {\rm\quad if \quad} 0.33 < n_B(\omega_0) <
300\\
0.5\cdot 10^{-3}  & {\rm\quad if \quad} n_B(\omega_0) < 0.33
\end{array}\right.\nonumber
\end{eqnarray}
The exact numerical solution of the rate equations
Eq.~(\ref{Eq:ME_Strong_Driving}) is shown in Figure~\ref{Fig:Cool}.

\subsection{Cooling by applying voltage driving}

Another way to achieve AC cooling is by applying radio frequency
voltage bias to the gates. In Fig.~\ref{Fig:SQUIDsys}, apply a
driving voltage $V_{x}=V_{0}\cos \omega_{\rm d}\,t$ on the resonator
and another driving voltage $V_{g}=-(C_{x}/C_{g})V_{x}$ on the
CPB. The ac voltage $V_{x}$ generates resonant coupling between
the mechanical resonator and the CPB when $\omega
_{ac}=E_{J}-\omega_{0}$, which corresponds to the first red
sideband coupling in quantum optics. The voltage $V_{x}$ also
generates an oscillating charge bias on the CPB with $\delta
N_{g}^{x}=C_{x}V_{x}/2e$; however, it is balanced by the bias
$V_g$, which prevents harmful ac pumping of the CPB.

{\it The scheme.} The Hamiltonian of the voltage driving scheme is
\begin{eqnarray}
H &=&-\frac{E_{J}}{2}\sigma_{z}+\hbar \omega_{0}\hat{a}^{\dagger }\hat{a}%
+4E_{c}\delta n_{g}\sigma _{x}  \nonumber \\
&&+\lambda \cos \omega_{\rm d}\,t\left( \hat{a}^{\dagger }+\hat{a}\right) \sigma
_{x}-\frac{\hat{X}}{2}\sigma _{x}
\end{eqnarray}%
where the qubit works at the optimal point of $\delta n_{g}=0$ to
avoid the $1/f$ noise. The coupling is
\begin{equation}
\lambda =-4E_{c}\frac{C_{x}V_{0}}{2e}\frac{\Delta x}{d}.  \label{lambda}
\end{equation}%
Similar to the ac flux driving setup, we analyze the sideband
cooling  ($\omega_0\gg \lambda, \Gamma _{r}$) in the regimes of
weak driving and the strong driving; both can be reached in
experiments.

The analysis is essentially the same as that in the flux driving
case. Instead of the second order matrix element $\Delta$ (see
Eq.~(\ref{Eq:Delta})), we have the direct coupling $\lambda/2$.
Thus to obtain the cooling and the heating rates we substitute
$\Delta \rightarrow \lambda/2$ into Eqs.~(\ref{Eq:Cooling_Rates})
and (\ref{Eq:Heating_Rates}). In this scheme there is no driving
induced contribution of the low frequency ($1/f$) noise similar to
(\ref{Eq:Heating_Rates_1/f}).

{\it Strong driving.} The strong driving regime is achieved for
$\lambda \sqrt{n}/2\gg \Gamma _{r}/2$. As the $1/f$ noise does not
contribute to the heating, i.e.,
the rates (\ref{Eq:Heating_Rates_1/f}) do not appear,
the leading heating mechanism is the off-resonance coupling (see
Eq.~(\ref{Eq:Heating_Rates})). Thus, the dynamics is described by
the rate equation (\ref{Eq:ME_Strong_Driving}) with $\Gamma_{\rm
e}=0$. We obtain the cooling results as follows: For $n_{B} >
Q(\Gamma_{r}/2\omega_0)$ we obtain $\left\langle n\right\rangle
\approx n_{B} - Q(\Gamma_{r}/2\omega_0)$. In the
intermediate regime
$Q(\Gamma_{r}/2\omega_0)>n_{B}>Q(\Gamma_{r}/\omega_0)(\lambda/4\omega_0)^2$
the result is $\left\langle n\right\rangle\approx \omega
_{0}n_{B}/Q\Gamma_{r}$. Finally, for
$n_{B}<Q(\Gamma_{r}/\omega_0)(\lambda/4\omega_0)^2$, the
occupation number saturates at $\left\langle n\right\rangle
\approx (\lambda/4\omega_0)^2$.

For example, let $\Gamma _{r}=5\cdot 10^{-3}{\rm \mu eV}$, $%
\lambda =25\cdot 10^{-3}{\rm \mu eV}$ (with the voltage bias $V_{0}\sim 50\,%
{\rm mV}$ ) and $Q=10^{5}$. We have $(\lambda /4
\omega_{0})^2\Gamma _{r} \approx 8 \cdot 10^{-7}{\rm \mu eV}$. The
stationary occupation number is
\begin{eqnarray}
\left\langle n\right\rangle \approx \left\{%
\begin{array}{cl}
n_{B}\left( \omega _{0}\right)-500 &{\rm\quad if \quad} n_{B}(\omega_0)> 500 \\
[1mm]
10^{-3} n_{B}(\omega_0) &{\rm\quad if \quad} 0.15 < n_{B}(\omega _{0}) < 500 \\
[1mm] 1.5 \cdot 10^{-4} &{\rm\quad if \quad} n_{B}(\omega _{0}) <0.15 
\end{array}\right.\nonumber
\end{eqnarray}%
which gives $\left\langle n\right\rangle =2.5\cdot 10^{-3}$ at
the temperature of $20\,{\rm mK}$.

{\it Weak driving.} For $\lambda \sqrt{n}/2\ll \Gamma _{r}/2$, the
dynamics of the resonator is described by
the rate equation (\ref{Eq:ME_Weak_Driving}) with
$A_{-}=\lambda^2/\Gamma_{r}$ and
$A_{+}=(\lambda/4\omega_{0})^2\Gamma_r$. Note that the low frequency
noise doesn't appear in this scheme ($\Gamma_{\rm e}=0$),
which improves the cooling efficiency.
For $n_{B}(\omega_{0}) < A_{+}Q /
\omega _{0}$ the average occupation is given by
$\left\langle n\right\rangle = A_{+}/A_{-}$; for $n_{B}(\omega_{0}) > A_{+}Q / \omega _{0}$ we obtain
$\left\langle n\right\rangle=n_{B}(\omega_{0})\omega _{0}/A_{-}Q$. As an example, let $%
\Gamma _{r}=50\cdot 10^{-3}{\rm \mu eV}$, $\lambda =5\cdot 10^{-3}{\rm \mu eV%
}$ (with the bias voltage of $V_{0}\sim 10\,{\rm mV}$ ) and $Q=10^{5}$. We have $%
A_{-}=0.5\cdot 10^{-3}{\rm \mu eV}$ and $A_{+}=3\cdot 10^{-7}{\rm \mu eV}$.
Then, the cooling results are as follows:
\begin{equation}
\left\langle n\right\rangle =\left\{%
\begin{array}{cc}
10^{-2}n_{B}(\omega _{0}), & n_{B}(\omega _{0}) >0.06 \\ [1mm]
6\cdot 10^{-4}, & n_{B}(\omega _{0}) <0.06%
\end{array}\right.
\label{n_st}
\end{equation}%
where at the temperature of $20\,{\rm mK}$ with
$n_{B}(\omega_{0})=5$, we have $\left\langle n\right\rangle=0.05$.
Thus better cooling can be achieved in the strong driving regime
than that in the weak driving regime.

{\it Discussion.}  The relaxation rate $\Gamma _{r}$ can be
adjusted by varying the external circuit of the CPB. For example,
by varying the gate capacitance $C_{g}$, the relaxation rate changes as $\Gamma_r\propto%
C_{g}^{2}$. The coupling constant $\lambda $ can be adjusted by
varying the bias $V_{0}$. In this scheme, we choose $V_{0}$ to be
in the range of $10\ldots 100\,{\rm mV}$ and $\Gamma _{r}$ in the
range of $(1\ldots 100)\cdot 10^{-3}{\rm \mu eV}$, which includes
both the weak driving regime $\left( \lambda \sqrt{n}/2\ll \Gamma
_{r}/2\right) $ and the strong driving regime $\left( \lambda
\sqrt{n}/2\gg \Gamma _{r}/2\right)$.  These parameter regimes have
been realized in charge qubit experiments.

One practical issue of this scheme concerns the accuracy of the
gate compensation. Both gate voltages generate an extra part of
the CPB's charging energy $\delta H=4E_{c}\delta n_{g}\sigma
_{x}$, with $\delta n_{g}=\left (C_{x}V_{x}+C_{g}V_{g}\right
)/2e$. By controlling the voltage with an accuracy of microvolts,
which can be achieved with standard technology, the oscillating
bias on the CPB can be neglected.

Compared with the ac flux driving cooling scheme in this paper,
the resonator can now be cooled to a lower temperature because
the heating process only involves the electromagnetic noise of the circuit at
frequency $E_{J}$, while the low frequency noise, which is the dominant
heating factor in the previous scheme, doesn't affect the system.

\section{DC cooling}
\label{sec:dc} 
\begin{figure}
\includegraphics[width=0.55\columnwidth]{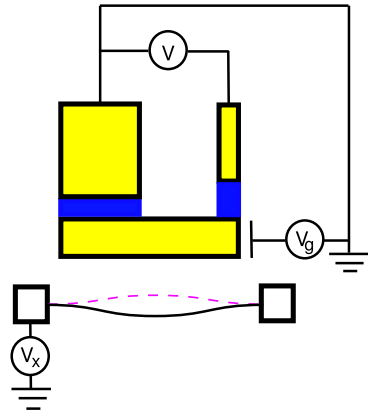}
\caption{The system with DC driving.} \label{Fig:Vsys}
\end{figure}

Effectively, AC driving can be achieved applying a
DC transport voltage to an auxiliary Josephson junction. We modify
the system as shown in Fig.~\ref{Fig:Vsys} so that it becomes
effectively an SET transistor. Dissipative Cooper pair and
quasi-particle transport is similar systems was considered in
Refs.~\cite{Averin89,Geerligs}.
The Hamiltonian in the charge basis reads
\begin{eqnarray}
\label{Eq:Charge_Hamiltonian_V} H &=& \frac{(Q-C_{\rm R}V-C_{\rm
g}V_{\rm g}-C_x(x)V_x)^2}{2C_{\Sigma}}\nonumber \\&-& E_{\rm
J,L}\cos\theta - E_{\rm J,R}\cos(\theta + 2eVt/\hbar) + H_x \ ,
\end{eqnarray}
where $\theta$ is the phase on the island (and simultaneously on
the left junction as the left lead is grounded), $C_{\Sigma}\equiv
C_{\rm L} + C_{\rm R} + C_{\rm g} + C_x(x)$. Due to the transport
voltage $V$, the Hamiltonian is time dependent and, thus,
represents a driven system. We repeat the steps described above
and arrive (in the charge basis) at
\begin{eqnarray}
\label{Eq:Spin_Hamiltonian_V} H =&-& \frac{E_{\rm
J,L}}{2}\,\sigma_x - \frac{X}{2}\, \sigma_z + \hbar \omega_0
a^{\dag}a + \lambda \,(a^{\dag} + a)\,\sigma_z \nonumber \\&-&
\frac{E_{\rm J,R}}{2}(\sigma_+\,e^{i\omega_{\rm J}t+i\delta\phi} + {\rm h.c.})
\ ,
\end{eqnarray}
where $X \equiv -\frac{4 E_{\rm C}}{e}\,\left(C_{\rm R}\delta V +
C_{\rm g} \delta V_{\rm g} + C_x \delta V_x \right)$, $\delta
\phi \equiv (2e/\hbar)\int \delta V \, dt$, and the Josephson frequency
$\omega_{\rm J} \equiv 2eV/\hbar$.
We see that the right junction's
Josephson energy assumes the role of pumping amplitude $\Omega$,
while $\omega_{\rm J}$ is the pumping frequency. Unfortunately $V$ is noisy
and, thus, the pumping source has a substantial line-width. The
transport voltage should not be sensitive to the background
charges, therefore $V$ is assumed to have Ohmic noise spectrum (no
$1/f$ component).

After a $\pi/2$ rotation in the $x-z$ plane, $\sigma_x
\leftrightarrow \sigma_z$, we obtain
\begin{eqnarray}
\label{Eq:Spin_Hamiltonian_V_rotated} H =&-& \frac{E_{\rm
J,L}}{2}\,\sigma_z - \frac{X}{2}\, \sigma_x + \hbar \omega_0
a^{\dag}a + \lambda \,(a^{\dag} + a)\,\sigma_x \nonumber \\&-&
\frac{E_{\rm J,R}}{4}\left((\sigma_z + i\sigma_y)
\,e^{i\omega_{\rm J}t+i\delta\phi} + {\rm h.c.}\right)  \ .
\end{eqnarray}

While the Hamiltonians (\ref{Eq:Spin_Hamiltonian_pumped}) and
(\ref{Eq:Spin_Hamiltonian_V_rotated}) look similar, there are two important
differences. One, already discussed, is the fact that the pumping frequency $\omega_{\rm J}$
in (\ref{Eq:Spin_Hamiltonian_V_rotated}) is fundamentally noisy, while $\omega_{\rm d}$
in (\ref{Eq:Spin_Hamiltonian_pumped}) can be made coherent. The second (very
important) difference is that in (\ref{Eq:Spin_Hamiltonian_pumped}) the pumping
is applied to $\sigma_z$ only, while in (\ref{Eq:Spin_Hamiltonian_V_rotated})
it couples to $\sigma_z$ and $\sigma_y$. Both these facts hinder the cooling.
Indeed, the coupling to $\sigma_y$ gives a direct matrix element $E_{\rm
J,R}/4$ between the states $\ket{\uparrow,n,m}$ and $\ket{\downarrow,n,m+1}$.
This interaction repels the levels and we must choose $E_{\rm J,R} \ll
4\omega_0$ so that the resonant detuning as in Fig.~\ref{Fig:Stair_Cooling} is
possible. In addition, the noise of the transport voltage translates into the
line width for the transition $\ket{\uparrow,n,m} \rightarrow
\ket{\downarrow,n-1,m+1}$ equal to $\Gamma_\varphi = 2\pi \alpha_{\rm tr} k_{\rm B}T/\hbar$,
where $\alpha_{\rm tr} \equiv R/R_{\rm Q}$. The fluctuations of the transport
voltage are not screened by the ratio of capacitances as it happens for the
gate charge. Therefore $\alpha_{\rm tr} \approx 10^{-2}$. Because of these
additional constraints the applicability of the DC cooling scheme is limited to
higher frequency/quality factor resonators.  For an estimate, consider an
oscillator with $\omega_0 = 2\pi\times 1{\ \rm GHz} \approx 5\ \mu{\rm
eV}\approx 50$ mK at temperature $T = 50$ mK. We then obtain $\Gamma_\varphi
\approx 0.3 \ \mu{\rm eV}$, which significantly exceeds  $\Gamma_r$. Hence, we
have to substitute $\Gamma_r$ by
$\Gamma_\varphi$ in all formulas. For the Josephson coupling in the right
junction we take $E_{\rm J,R}=2\ \mu$eV.  Then, instead of
Eq.~(\ref{Eq:Delta}), we find $\Delta \approx E_{\rm J,R}\lambda/(2E_{\rm J,L})
\approx 2\cdot 10^{-3}\ \mu $eV (we assume $E_{\rm J,L} \approx 50\ \mu$eV).
The cooling rate can again be represented as $A_- n$, where $A_-\approx
2\Delta^2/\Gamma_\varphi \approx 2\cdot 10^{-5}\mu$eV. Thus, cooling becomes
possible only if $Q > \omega_0/A_- \approx 2.5\cdot 10^{5}$.

\section{Discussion and comparison with quantum optics.}

For comparison with quantum optics cooling schemes, we present here an analysis
of some of the processes described above using the quantum optics language. In
quantum optics literature one usually employs the transformation to the
rotating frame and/or other canonical transformations together with the
Rotating Wave Approximation (RWA) in order to single out the near resonant
terms responsible for the studied transitions. We start with Hamiltonian
(\ref{Eq:Spin_Hamiltonian_pumped}) and transform it into the interaction
representation with respect to $H_0(t) \equiv -[E_{\rm J}+\Omega+\Omega\cos
2\omega_{\rm d}\,t]\,\sigma_z/2 + \omega_0 a^{\dag}a$. This amounts to
\begin{equation}
\sigma_{+}\rightarrow \sigma_{+} e^{-i(E_{\rm J}t+\Omega t +
\frac{\Omega}{2\omega_{\rm d}}\sin 2\omega_{\rm d}\, t)}\ \ \ {\rm and} \ \
a\,\rightarrow a e^{-i\omega_0 t} \ .
\end{equation}
As $\Omega \ll 2\omega_{\rm d} \sim E_{\rm J}$ we can expand the factor
$\exp(-i(\Omega/2\omega_{\rm d})\sin 2\omega_{\rm d}\,t)$.
The near resonant (RWA) parts of the coupling term $H_{\lambda}
\equiv \lambda(a^{\dag}+a)\sigma_x$ read, then, after the
transformation as
\begin{equation}
\label{Eq:H_lambda_RWA}
H_{\rm \lambda}^{\rm RWA} = -\frac{\Omega\lambda}{4\omega_{\rm d}}\, (a
\sigma_{-} + a^{\dag}\sigma_{-}\,e^{2\omega_0 t} + h.c.) \ ,
\end{equation}
where the resonance condition $2\omega_{\rm d} = E_{\rm J} + \Omega - \omega_0$ was
assumed. The first term of (\ref{Eq:H_lambda_RWA}) clearly corresponds to the
resonant transition provided by the matrix element $\Delta \sqrt{n}$ (see
Eq.~(\ref{Eq:Delta})).  The second term corresponds to the off-resonant
transition. In analogy with atom optics~\cite{Leibfried_RMP}, we can introduce
Lamb-Dicke parameter $\eta \equiv \lambda/2\omega_{\rm d} \approx \lambda/E_{\rm J}$.
Note that, unlike the case of trapped atoms/ions we do not have near resonant
terms flipping just the spin, $\sim \Omega \sigma_{\pm}\exp(\pm i\omega_0)$.
Such an interaction would repel the levels, and, for a resonant detuning, one
would need $\Omega \ll \omega_0$. In our case, a much stronger driving is
allowed.  We do, however, encounter this limitation in the DC cooling scheme
due to the presence of the near-resonant term in the Hamiltonian
Eq.~(\ref{Eq:Spin_Hamiltonian_V_rotated}) which is proportional to $\sigma_y$.

Also, in the dissipative term $H_X \equiv -(X/2)\sigma_x$, we find a slow
contribution:
\begin{equation}
H_X^{\rm slow} = \frac{X}{2}\,\frac{\Omega}{4\omega_{\rm d}}\,\left(
\sigma_{-} e^{i\omega_0 t} + h.c.\right)
\ ,
\end{equation}
which is responsible for the heating process (\ref{Eq:Heating_Rates}) dominated by the low frequency noise.
In quantum optics this contribution is typically neglected as there are no
strong low frequency sources.

In the case of ac voltage cooling, in the rotating wave
approximation the Hamiltonian is
\begin{eqnarray}
H^{RWA} &=&-\frac{\hbar \omega _{0}}{2}\sigma _{z}+\hbar \omega _{0}\hat{a}%
^{\dagger }\hat{a}  \nonumber \\
&&+\frac{\lambda }{2}\left( \hat{a}^{\dagger }\sigma _{+}+\hat{a}\sigma _{-}+%
\hat{a}^{\dagger }\sigma _{-}+\hat{a}\sigma _{+}\right)  \label{Hrwa} \\
&&-\frac{\hat{X}}{2}\left( \sigma _{+}e^{-i\omega_{\rm d}\,t}+
\sigma _{-}e^{i\omega_{\rm d}\,t}\right)  \nonumber
\end{eqnarray}%
where the interaction includes the resonant coupling
$\hat{a}^{\dagger }\sigma _{+}+\hat{a}\sigma _{-}$ between the
states $\left| \uparrow ,n\right\rangle $ and $\left|
\downarrow,n-1\right\rangle $, and the off resonant coupling
$\hat{a}^{\dagger }\sigma _{-}+\hat{a}\sigma _{+}$ between the
states $\left| \uparrow ,n\right\rangle $ and $\left| \downarrow
,n+1\right\rangle $ with an energy difference of $2\hbar \omega
_{0}$. The time dependence in the last term shows that only high
frequency fluctuations on the order of $E_{J}$ induces relaxation.

\begin{figure}
\includegraphics[width=0.75\columnwidth]{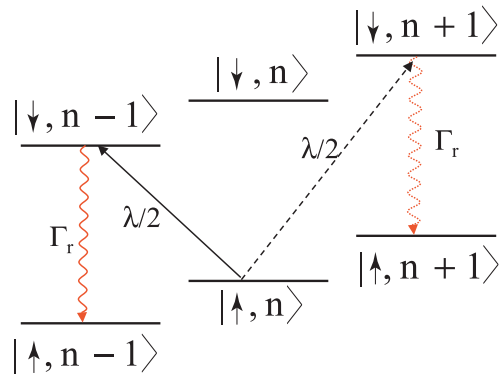}
\caption{Level structure of the voltage driving scheme for ac
cooling. The solid lines indicate the cooling process; and the
dotted lines indicate the heating process.}
\label{Fig:Voltage_Driving}
\end{figure}

The voltage driving scheme for ac cooling presents a direct
analogue to the laser cooling of quantum optics systems as is
obvious from Fig.~\ref{Fig:Voltage_Driving}; however, it is a
simpler scheme as the coupling is a direct bilinear coupling
instead of the polaron coupling in quantum optics.

\section{``Doppler'' cooling}

For smaller oscillator frequencies $\omega_0$ and/or larger qubit relaxation
rates $\Gamma_{\rm r}$, it may happen that the vibrational sidebands are no
longer resolved, i.e., $\Gamma_{\rm r} > \omega_0$. Then, cooling to the ground
state is impossible but the less effective ``Doppler'' cooling is still
feasible. In the standard Doppler cooling, cold atoms moving in a laser field
experience different light pressure depending on their velocity. For instance,
if the laser is red-detuned relative to the atomic transition, then atom moving
towards the light source will be absorbing photons more than an atom moving
away from the source, which will result in a velocity-dependent force on the
atom.  By properly arranging multiple lasers it is then possible to achieve
significant reduction of the atom's effective temperature. The same reasoning
applies to an atom in a trap, when the trap frequency is much smaller than the
atomic transition line width.  Equivalently, in this case, the Doppler cooling
can be reinterpreted in terms of the cooling and heating transition rates for
the oscillating atom.  For red-detuned laser, photon absorption processes with
simultaneous phonon emission dominate the ones where an additional phonon is
created.  One can show that, similar to the free atom case, this leads to
cooling down to temperatures proportional to the atomic transition line width.
In this section we will demonstrate that an analog of Doppler cooling can be
performed in both AC and DC cooling setups described in the previous sections
in the regime of non-resolved sidebands.

{\it AC cooling scheme.} We start with the AC-driving scheme described by
Hamiltonian (\ref{Eq:Spin_Hamiltonian_pumped}). In the Doppler case it is
enough to study, first, the spin's dynamics without the oscillator. We perform
the following transformation
\begin{equation}
\sigma_{+}\rightarrow \sigma_{+} e^{-i(2\omega_{\rm d}\,t +
\frac{\Omega}{2\omega_{\rm d}}\sin 2\omega_{\rm d}t)}
\ ,
\end{equation}
and keep only the RWA terms in the coherent part of the Hamiltonian but all the
terms in the part describing the interaction with the bath. The transformed
Hamiltonian reads
\begin{eqnarray}
\label{Eq:AC_Doppler_Hamiltonian} H_{\rm spin}^{\rm RWA} = \frac{\delom}{2}\,
\sigma_z - \frac{X}{2}\,
(\sigma_{+}e^{-2i\omega_{\rm d} t} + h.c.) +
\frac{\Omega}{4\omega_{\rm d}}\,\frac{X}{2}\,\sigma_x
\ ,
\nonumber \\
\end{eqnarray}
where $\delom$ is the detuning $\delom \equiv 2\omega_{\rm d}-E_{\rm J} -\Omega$. The
last term in the Hamiltonian, as before (see
Eqs.~(\ref{Eq:Qubit_DI_Excitation}) and (\ref{Eq:Heating_Rates})), generates
transitions between the ground and excited states of the qubit. The
corresponding rate, however, is typically much smaller than the relaxation rate
generated by the second term of the Hamiltonian, and hence can be neglected.
This is in contrast to the AC ground states cooling scheme discussed above,
where the qubit heating rate determined the lowest possible occupation number
of the resonator.   The slow (RWA) part of the interaction Hamiltonian reads
\begin{equation}
\label{Eq:AC_Doppler_Interaction}
H_{\lambda}^{\rm slow} =
-\frac{\Omega}{4\omega_{\rm d}}\,\lambda\,(a^{\dag}+a)\,\sigma_x
\ .
\end{equation}
With Hamiltonian (\ref{Eq:AC_Doppler_Hamiltonian}) we write down the
Bloch-Redfield equations~\cite{Bloch_Derivation,Redfield_Derivation} for the
spin's density matrix $\dot \rho = R\rho$, where $R$ is the Redfield tensor. In
this equation, the density matrix is treated as a four-vector. We choose the
representation
$\rho = (1/2) s \sigma_0 + \gamma \sigma_z + \alpha \sigma_{+} + \alpha^{*}\sigma_{-}$, which
gives $\rho=(s,\gamma,\alpha,\alpha^*)$ and ${\rm Tr}\rho = s$. For proper
density matrices $s=1$. In this representation
\begin{equation}
\label{Eq:AC_Redfield}
R = \left(
\begin{array}{cccc}
0 & 0 & 0 & 0\\
\frac{\Gamma_{\rm r}}{2} & -\Gamma_{\rm r} & 0 & 0\\
0 & 0 & -i\delom  -\frac{\Gamma_{\rm r}}{2} & 0 \\
0 & 0 & 0 & i\delom -\frac{\Gamma_{\rm r}}{2}
\end{array}
\right)\ .
\end{equation}
The rate $\Gamma_{\rm r}$ originates from the second term of
(\ref{Eq:AC_Doppler_Hamiltonian}),
\begin{equation}
\label{Eq:Doppler_Gamma_r}
\Gamma_{\rm r} =
\frac{\langle X^2_{\nuom=2\omega_{\rm d}}\rangle}{4\hbar^2}
\ ,
\end{equation}
(cf. Eq.~\ref{Eq:Gamma_r}).  For simplicity we have assumed that
at high frequency $2\omega_{\rm d}$ the temperature is effectively
zero, i.e., $\langle X^2_{\nuom=-2\omega_{\rm d}}\rangle \ll
\langle X^2_{\nuom=2\omega_{\rm d}}\rangle$.

Next we employ the ``quantum regression theorem'' (approximation) to obtain
(for $t>0$) the correlator
\begin{equation}
C_x(t) \equiv \langle \sigma_x(t)\sigma_x(0)\rangle = {\rm Tr}\left(\sigma_x\,
e^{Rt}\,\sigma_x\,\rho_{\infty}\right)
\ ,
\end{equation}
where $\rho_{\infty}$ is the stationary density matrix, $R\rho_{\infty}=0$. For
$t<0$ we can use $C_x(-t) = C_{x}^{*}(t)$. In the four-vector representation
the operator $\sigma_x$ multiplying from the left is given by
\begin{equation}
\label{Eq:sigma_x}
\sigma_x = \left(
\begin{array}{cccc}
0 & 0 & 1 & 1\\
0 & 0 & -1/2 & 1/2\\
1/2 & -1 & 0 & 0 \\
1/2 & 1 & 0 & 0
\end{array}
\right)\ .
\end{equation}
Finally, after the Fourier transform we obtain
\begin{equation}
C_x(\nuom) = - 2 {\rm Re}\, {\rm
Tr}\left(\sigma_x\,[R+i\nuom]^{-1}\,\sigma_x\,\rho_{\infty}\right)
\ ,
\end{equation}
which is easy to calculate (using Mathematica). Expanding near $\nuom=0$ we
obtain
$C_x(\nuom) = S_x + \eta_x\nuom$, where
\begin{equation}
S_x = \frac{4\Gamma_{\rm r}} {4\delom^2 + \Gamma^2_{\rm r}}
\ ,
\end{equation}
and
\begin{equation}
\eta_x = -\frac{32\delom\,\Gamma_{\rm r}}
{(4\delom^2 + \Gamma_{\rm r}^2)^2}
\ .
\end{equation}
From the spin correlation functions, we can now determine the
transition rates for the resonator, $A_\pm =
(\Omega\lambda/4\omega)^2C_x(\mp \omega_0)$. Note, that the same
expressions for $A_\pm$ can be obtained from
Eq.~(\ref{Eq:Golden_Rule}) generalized to arbitrary detuning.
Then, the secondary ``heat bath'' introduced by cooling is
characterized by the effective temperature
\begin{equation}
T^* \equiv \frac{S_x}{2\eta_x} = -\frac{4\delom^2 + \Gamma_{\rm
r}^2} {16\delom} \ .
\end{equation}
Optimizing with respect to the detuning $\delom$, we find that the minimum
(positive) temperature $T_{\rm min} \approx \Gamma_{\rm r}/4$ is reached for the optimal
red detuning $\delom_{\rm opt}\approx -\Gamma_{\rm r}/2$. For the effective quality
factor due to the spin we obtain
\begin{equation}
\label{Eq:Doppler_Qeff_AC}
{1 \over Q^*} = {A_- - A_+ \over \omega_0} = \frac{\Omega^2\lambda^2
\eta_x}{8\omega_{\rm d}^2}
\ .
\end{equation}
The final resonator occupancy can now be obtained from Eq.~(\ref{Eq:n_2baths}).
Clearly, $Q^{*}$ must be smaller than the oscillator's own quality factor in
order for cooling to be effective.  For the same parameters as used in the AC
cooling section, we find that at the optimal detuning $Q^*\sim 1$, which
corresponds to a nearly overdamped regime, similar to optical molasses in atom
optics.  Hence, a resonator with a frequency $\omega_0 \lesssim \Gamma_r/2 =
300$ kHz can be cooled down to temperature $T_{\rm min} \approx \Gamma_{\rm
r}/4 \sim 10\ \mu$K.  Note, that the effect of the dominant qubit heating
mechanism due to the $1/f$ noise, which we neglected here, if necessary, can be
managed by reducing the driving strength $\Omega$ (at the expense of reduced
cooling power, i.e. larger $Q^*$).

Similar analysis applies to the voltage driving scheme for AC
cooling in the Doppler regime. For $\lambda\sqrt{n}/2 \ll
\Gamma_{r}/2$, from Eq.~(\ref{Eq:Golden_Rule})
\begin{equation}
A_{-}=\frac{\lambda^2\Gamma _{r}}{\Gamma_{r}^2+4\left (
\delta\omega+\omega_{0} \right )^2}\  \quad
A_{+}=\frac{\lambda^2\Gamma _{r}}{\Gamma_{r}^2+4\left (
\delta\omega-\omega_{0} \right )^2}\ ,
\end{equation}
where $\delta\omega=\omega_{d}-E_{\rm J}$ is the detuning of the
driving frequency from the qubit's ground-to-excited-state
transition. Cooling is achieved when $\delta\omega < 0$.
Neglecting the effect of the finite intrinsic $Q$-factor, the
final phonon number from Eq.~(\ref{Eq:n_weak}) is
\begin{equation}
\langle n \rangle_f = \frac{\Gamma_r^2+4\left (
\delta\omega+\omega_{0}\right )^2}{16|\delta\omega|\omega_{0}}
\end{equation}
The scheme provides optimal cooling at $\delta\omega_{\rm
opt}=-\Gamma_{r} / 2$ with the temperature $T_{\rm min} =
\Gamma_{r}/4$, same as the flux driving case.

{\it DC cooling scheme}. Analysis of the DC cooling scheme
proceeds analogously.  We start with Hamiltonian
(\ref{Eq:Spin_Hamiltonian_V_rotated}) and perform transformation
\begin{equation}
\sigma_{+}\rightarrow \sigma_{+} e^{-i(\omega_{\rm J} t +
\frac{\Omega}{\omega_{\rm J}}\sin \omega_{\rm J} t)}
\ ,
\end{equation}
where, for now, $\Omega\equiv E_{\rm J,R}$, and we remind that $\omega_{\rm J} \equiv (2e/\hbar)V$.
For simplicity we consider $V$
noiseless now but later will introduce the low-frequency (classical) noise of
$V$. Then, we obtain
\begin{eqnarray}
\label{Eq:DC_Doppler_Hamiltonian} H_{\rm spin}^{\rm RWA} &=& \frac{\delom}{2}\,\sigma_z
-\frac{\Omega}{4}\,\sigma_x \nonumber \\
&-&
\frac{X}{2}\,(\sigma_{+}e^{-V t} + h.c.) + \frac{\Omega}{2V}\,\frac{X}{2}\,\sigma_x
\ ,
\end{eqnarray}
where $\delom \equiv \omega_{\rm J}-E_{\rm J,L}$. The main differences with
(\ref{Eq:AC_Doppler_Hamiltonian}) are: i) the second term of
(\ref{Eq:DC_Doppler_Hamiltonian}) is absent in
(\ref{Eq:AC_Doppler_Hamiltonian}) (recall the extra matrix element in the DC
scheme); ii) the detuning $\delom$ is noisy due to the noise of $V$. This will
give an extra ``pure'' dephasing rate $\Gamma_\varphi = 2\pi \alpha_{\rm tr}
k_{\rm B}T/\hbar$, where $\alpha_{\rm tr} \equiv R/R_{\rm Q}$. As in the AC Doppler case,
neglecting the excitation rate coming from the last term in the Hamiltonian,
for the Redfield tensor we obtain
\begin{eqnarray}
\label{Eq:DC_Redfield}
R =
\left(\begin{array}{cccc}
0 & 0 & 0 & 0\\[1mm]
\frac{\Gamma_{\rm r}}{2} & -\Gamma_{\rm r} & -\frac{i\Omega}{4} &
\frac{i\Omega}{4}\\[1mm]
0 & -\frac{i\Omega}{2} & -i\delom  -\tilde{\Gamma}
& 0 \\[1mm]
0 & \frac{i\Omega}{2} & 0 & i\delom -\tilde{\Gamma}
\end{array}\right)
\end{eqnarray}
where $\tilde{\Gamma} = \Gamma_\varphi + \Gamma_{\rm r}/2$.  The general
expressions for $S_x$ and $\eta_x$ are quite complicated. An analysis shows
that the simplest cooling regime is achieved when $\Omega \ll
\Gamma_{\rm r} \approx \Gamma_\varphi$.  That is, it does not make sense to
keep small $\Gamma_{\rm r}$ when the line width is anyway given by large
$\Gamma_\varphi$. The relaxation rate $\Gamma_{\rm r}$ can easily be increased
by choosing bigger gate capacitances $C_x$ and/or $C_{\rm g}$. Assuming
$\Gamma_{\rm r} = \Gamma_\varphi$ and $\Omega \ll \Gamma_\varphi$ we obtain
\begin{equation}
S_x = \frac{12\,\Gamma_\varphi} {4\delom^2 + 9\,\Gamma_\varphi^2}
\ ,
\end{equation}
and
\begin{equation}
\eta_x = -\frac{96\delom\,\Gamma_\varphi}
{(4\delom^2 + 9\,\Gamma_\varphi^2)^2}
\ .
\end{equation}
Optimizing with respect to $\delom$, we find that the minimum temperature
$T_{\rm min} \approx (3/4)\Gamma_\varphi$ is achieved for the optimal detuning $\delom
\approx -(3/2)\Gamma_\varphi$.   The effective quality factor due
to the cooling environment, similarly to (\ref{Eq:Doppler_Qeff_AC}), is
\begin{equation}
\label{Eq:Doppler_Qeff_DC}
{1 \over Q^*} =
\frac{\Omega^2\lambda^2 \eta_x}{2\omega_{\rm J}^2}
\ .
\end{equation}
The regime $\Gamma_{\rm r} \ll \Gamma_\varphi$ is more subtle and
requires further analysis.

{\it Example:} Choosing $C_x \approx 200$ aF instead of previously assumed
$20$ aF, we obtain $\Gamma_{\rm r} \approx 0.3\ \mu{\rm eV} \approx
\Gamma_\varphi$.  We choose $\Omega = E_{\rm J,R} \approx 0.1\ \mu$eV (note that
this is quite a small value for usual Josephson junctions).  Then we obtain
$T_{\rm min}\approx 0.25\ \mu{\rm eV}\approx 2.5$ mK for $\delom \approx 0.5\
\mu{\rm eV} \approx 2\pi\times 100$ MHz.
Near optimal detuning point, we obtain $\eta_x \approx 6\, (\mu{\rm eV})^{-2}$.
For the coupling constant $\lambda$ we can take $\lambda \approx 1\ \mu$eV
instead of $0.1\ \mu$eV as we have allowed ten times bigger capacitance $C_x$.
Then we obtain $Q^* \approx 0.8\cdot 10^5$. Thus, for cooling to be effective,
resonator $Q$-factor should exceed $10^5$ and the final resonator temperature
is determined according to Eq.~(\ref{Eq:n_2baths}) with $T^* = 2.5 \ \mu$K.

\section{Conclusions}

We considered several approaches to active cooling of mechanical
resonators using a coupling to a superconducting Josephson qubit.
In the resolved vibrational sideband regime, when the qubit level
width is smaller than the resonator frequency, we proposed two
schemes for ground-state cooling of the resonator. In the first
scheme, the AC driving required for cooling is provided by an
external microwave source. We find that for a 100 MHz oscillator
coupled to a practically realizable Josephson qubit, at the
external temperatures below 1 K, it is possible to reduce the
thermal occupancy of the oscillator mode by three order of
magnitude. In the second scheme, the AC driving is generated by
the AC Josephson oscillations on an auxiliary junction of the
qubit.  This scheme is attractive since there is no need for an
external AC driving source; however, in the present realization,
we find that it is not as effective as the one with an explicitly
applied AC driving.  We also demonstrate that even in the regime
when the vibrational sidebands are not resolved, it is possible to
perform an analogue of Doppler cooling with the final resonator
temperature limited by the qubit line width.

\section{Acknowledgments}

We would like to acknowledge useful discussions with K. Schwab, D.
Mozyrsky, G. Sch\"on, S. Habib, Yu. Makhlin, G. Johansson. A.S.
thanks T-11 group of LANL for hospitality and acknowledges support
of CFN (DFG). This work was supported by the U.S. DoE. L.T. thanks
TFP group of Karlsruhe for hospitality. Work at the University of
Innsbruck is supported by the Austrian Science Foundation,
European Networks and the Institute for Quantum Information.


\end{document}